\documentclass[pra,aps,showpacs,twocolumn]{revtex4}
\usepackage{amsfonts}
\usepackage{amsmath}
\usepackage{amssymb}
\usepackage[dvips]{graphicx}

\begin{document}

\title{Abrupt Changes in the Dynamics of Quantum Disentanglement}
\date{\today}
\author{F. Lastra$^{1}$, G. Romero$^{1}$, C. E. L\'{o}pez$^{1}$, M. Fran\c{c}a
Santos$^{2}$, and J.C. Retamal$^{1}$}
\affiliation{$^{1}$Departamento de F\'{\i}sica, Universidad de
Santiago de Chile, USACH, Casilla
307 Correo 2, Santiago, Chile. \\
$^{2}$Departamento de F\'{\i}sica, Universidade Federal de Minas
Gerais, Belo Horizonte, Brazil.} 

\begin{abstract}
Entanglement evolution in high dimensional bipartite systems under
dissipation is studied. Discontinuities for the time derivative of
the lower bound of entanglement of formation is found depending on
the initial conditions for entangled states. This abrupt changes
along the evolution appears as precursors of entanglement sudden
death.
\end{abstract}

\maketitle

Entanglement is a cornerstone of modern quantum
physics~\cite{Schroedinger, nielsen}. The evolution of
entanglement in open quantum systems is a matter of increasing
interest and new phenomena have been predicted~\cite {Diosi, Dodd,
eberly00, eberly1,Marcelo,jamroz,ficek,Derkacz06}. One of the most
outstanding effects arises when entanglement vanishes long before
coherence is lost. It has been pointed out that systems composed
of two qubits in a noisy environment can lose its entanglement in
finite time, a phenomena named as Entanglement Sudden Death (ESD),
even though full decoherence happens asymptotically. This feature
appears for certain classes of states of two qubits under the
action of independent reservoirs. Examples of these classes are
the so-called \textquotedblleft X\textquotedblright mixed states
as well as some particular types of non-maximally entangled pure
states~\cite {Marcelo}.

The purpose of this work is to explore dynamical behavior of
entangled states in larger bipartite systems  under the action of
independent reservoirs. We show that unlike the case of two
qubits, $3\otimes 3$ systems may present not only ESD, but also
intermediate abrupt changes in the disentanglement dynamics, i.e.
the rate in which a given state loses its entanglement may change
throughout the dissipative process even though coherence is lost
in a constant rate. We show that these rate changes are associated
with sudden changes in the rank of the partially transposed
density matrix, which also provides an explanation for the sudden
death of entanglement.

We analyze the disentanglement dynamics for different initial
states and show that abrupt changes may be present or not
depending on the variation of a small number of parameters. We
also recover the result for two qubits when preparing the initial
state in a $2\otimes 2$ subspace of the whole system. Finally, we
interpret these results in terms of changes in the set of
entanglement witnesses appropriated for the characterization of
the entangled state in each part of the dynamics.

In this work we use a general measurement for the lower bound of
Entanglement of Formation (EOF) for a mixed state in $m\otimes n$
dimensions, which has been recently proposed~\cite{Albeverio}.
This proposal is based on the comparison between two major
criteria: $(i)$ the positivity under partial transposition (PPT
criterion)~\cite{Peres,Horodecki} and $(ii)$ the realignment
criterion~\cite{Rudolph,Chen}. EOF for $m\otimes n$-dimensional
systems ($m \le n$) is defined as~\cite{Albeverio}
\begin{widetext}
\begin{equation}
E(\rho )\geq \left\{
\begin{tabular}{ll}
$0$ & if $\Lambda =1,$ \\
$H_{2}\left[ \gamma (\Lambda )\right] +\left[ 1-\gamma (\Lambda
)\right]
\log _{2}(m-1)$ & if $\Lambda \in \left[ 1,\frac{4(m-1)}{m}\right] ,$ \\
$\frac{\log _{2}(m-1)}{m-2}(\Lambda -m)+\log _{2}(m)$ & if
$\Lambda \in \left[ \frac{4(m-1)}{m},m\right] ,$
\end{tabular}
\right.   \label{eof}
\end{equation}
\end{widetext}where $m$ is the dimension of the first subsystem and $\gamma $
is given by%
\begin{equation}
\gamma (\Lambda )=\frac{1}{m^{2}}\left[ \sqrt{\Lambda }+\sqrt{
(m-1)(m-\Lambda )}\right] ^{2},
\end{equation}%
with $\Lambda =\max (\left\Vert \rho ^{T_{A}}\right\Vert
,\left\Vert R(\rho )\right\Vert )$ and $H_{2}(x)=-x\log
(x)-(1-x)\log (1-x)$. The trace norm $\left\Vert \cdot
\right\Vert$ is defined by $\left\Vert G\right\Vert
=tr(GG^{\dagger })^{\frac{1}{2}}$. The matrix $\rho ^{T_{A}}$ is
the partial transpose with respect to the subsystem $A$, that is,
$\rho
_{ik,jl}^{T_{A}}=\rho _{jk,il}$, and the matrix $R(\rho )$ is defined as $%
R(\rho )_{ij,kl}=\rho _{ik,jl}$.

The PPT criterion says that $\rho ^{T_{A}}\geq 0$ for a separable
state~\cite{Peres}. On the other hand, the realignment criterion
says that a realigned version of $\rho $, for a separable state
must satisfy the condition: $\left\Vert R(\rho )\right\Vert \leq
1$. These conditions state that entanglement exists for $\Lambda
>1$. The maximum values that $\Lambda (t)$ can assume depend on
the dimensions of the bipartite systems. For example, for a
maximal two qutrits entangled state $\Lambda =3$, for two qubits
$\Lambda =2$. The minimum value for a separable state is always
$\Lambda =1$. In this work we use this quantity to study the time
evolution of entanglement in the presence of dissipation.

Let us consider entangled quantum states of two qutrits, with at most two
excitations, in the presence of dissipation at zero temperature. Such
situation can be conveniently described by the evolution equation:
\begin{equation}
\dot{\rho}=\sum_{1,2}\frac{\Gamma _{i}}{2}\left[ 2c_{i}\rho c_{i}^{\dagger
}-c_{i}^{\dagger }c_{i}\rho -\rho c_{i}^{\dagger }c_{i}\right] ,
\label{master}
\end{equation}%
where $c_{i},c_{i}^{\dagger }$ describes annihilation and creation
operators for bosonic modes and $\rho $ is a $3\otimes 3$ density
matrix in the computational basis $\{\mid 0\rangle ,\mid 1\rangle
,\mid 2\rangle \}\otimes \{\mid 0\rangle ,\mid 1\rangle ,\mid
2\rangle \}$ of both qutrits.

Let us consider at first glance a class of initially mixed states of two
qutrits which corresponds to a modification of a maximally entangled state
given as follows:
\begin{equation}
\begin{tabular}{l}
$\rho (0)=\frac{1}{3}\left(
\begin{array}{ccccccccc}
1 & 0 & 0 & 0 & \lambda & 0 & 0 & 0 & \lambda \\
0 & 0 & 0 & 0 & 0 & 0 & 0 & 0 & 0 \\
0 & 0 & 0 & 0 & 0 & 0 & 0 & 0 & 0 \\
0 & 0 & 0 & 0 & 0 & 0 & 0 & 0 & 0 \\
\lambda & 0 & 0 & 0 & 1 & 0 & 0 & 0 & \lambda \\
0 & 0 & 0 & 0 & 0 & 0 & 0 & 0 & 0 \\
0 & 0 & 0 & 0 & 0 & 0 & 0 & 0 & 0 \\
0 & 0 & 0 & 0 & 0 & 0 & 0 & 0 & 0 \\
\lambda & 0 & 0 & 0 & \lambda & 0 & 0 & 0 & 1%
\end{array}
\right),$%
\end{tabular}
\label{initial}
\end{equation}
where $\lambda $ is a real parameter ranging from $0<\lambda <1$.
In the extreme cases, $\lambda =0,$ we have a separable state
whereas for $\lambda =1$ we have a maximally entangled state. The
Eq.(\ref{master}) can be solved for arbitrary decay constants, but
for simplicity we reduce the problem to the simplest case $\Gamma
_{1}=\Gamma _{2}=\Gamma $. By a numerical calculation we realize
that $\left\Vert \rho ^{T_{A}}\right\Vert \geq \left\Vert R(\rho
)\right\Vert $ for all times, so that, we need to concentrate only
in $\Lambda (t)=\left\Vert \rho ^{T_{A}}\right\Vert$.

\begin{figure}[t]
\includegraphics[width=60mm]{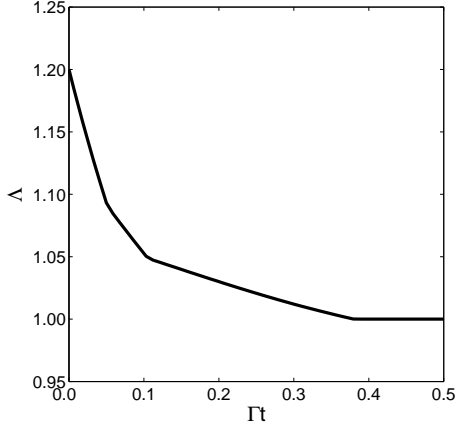}
\caption{{Evolution of $\Lambda (t)$ for the initial mixed state
of Eq.( \protect\ref{initial}) with $\protect\lambda =0.1$ as a
function of the adimensional time $\Gamma t$. }} \label{Fig1}
\end{figure}

\begin{figure}[t]
\includegraphics[width=60mm]{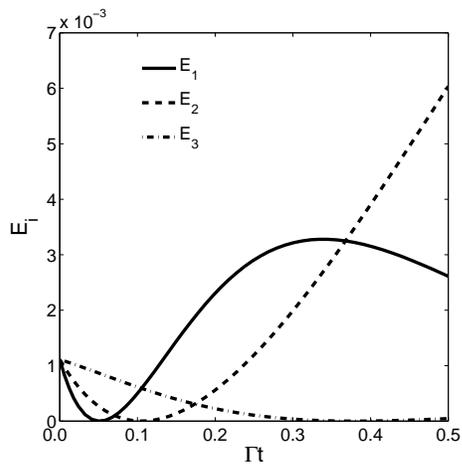}
\caption{{Relevant eigenvalues of matrix $M$ for the case shown in Fig.~%
\protect\ref{Fig1}.}}
\label{Fig2}
\end{figure}

\begin{figure}[t]
\includegraphics[width=60mm]{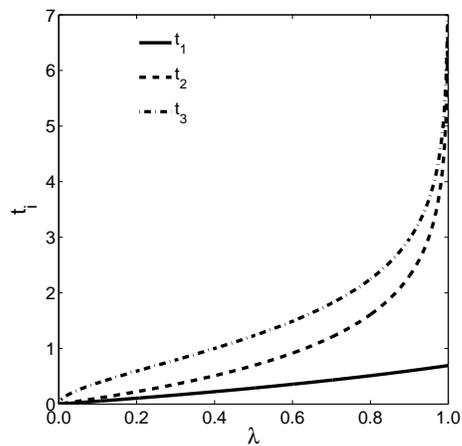}
\caption{{Times for sudden changes in the dynamics of entanglement as a
function of the parameter $\protect\lambda $. Solid line corresponds to the
first sudden change, dashed line corresponds to the second sudden change,
and dot-dashed line corresponds to the sudden death.}}
\label{Fig3}
\end{figure}

Fig.~\ref{Fig1} shows the evolution of $\Lambda (t)$ for $\lambda
=0.1$. As we observe, $\Lambda (t)$ undergoes sudden changes along
its evolution exhibiting discontinuous derivatives, finally
evolving to a situation where entanglement abruptly dies. As
compared with the case of two qubits a reacher dynamical behavior
of entanglement appears. From the definition of $ \Lambda (t)$,
this feature must be closely connected with the temporal
dependence of the eigenvalues of $M=\rho ^{T_{A}}\cdot \left( \rho
^{T_{A}}\right) ^{\dag }$. In our case both analytical and
numerical calculations of the eigenvalues of the matrix $M$ can be
carried out. From numerical calculations we realize that the
abrupt changes of entanglement evolution are dominated by the
behavior of a restricted number of eigenvalues given by:
\begin{equation}
\begin{tabular}{l}
$E_{1}\left( t\right) =\left( \rho _{12,12}\right) ^{2}+\left( \rho
_{11,22}\right) ^{2}-2\rho _{12,12}\rho _{11,22},$ \\
$E_{2}\left( t\right) =\left( \rho _{00,11}\right) ^{2}+\left( \rho
_{01,01}\right) ^{2}-2\rho _{00,11}\rho _{01,01},$ \\
$E_{3}\left( t\right) =\left( \rho _{00,22}\right) ^{2}+\left( \rho
_{02,02}\right) ^{2}-2\rho _{00,22}\rho _{02,02},$%
\end{tabular}
\label{eig}
\end{equation}%
where
\begin{equation}
\begin{tabular}{l}
$\rho _{12,12}=\frac{2}{3}\left( e^{-3\Gamma t}-e^{-4\Gamma t}\right) $ \\
\\
$\rho _{11,22}=\frac{\lambda }{3}e^{-3\Gamma t}$ \\
\\
$\rho _{00,11}=\frac{2\lambda }{3}e^{-3\Gamma t}-\frac{4\lambda }{3}%
e^{-2\Gamma t}+\lambda e^{-\Gamma t}$ \\
\\
$\rho _{01,01}=2e^{-3\Gamma t}-\frac{7}{3}e^{-2\Gamma t}-\frac{2}{3}%
e^{-4\Gamma t}+e^{-\Gamma t}$ \\
\\
$\rho _{00,22}=\frac{\lambda }{3}e^{-2\Gamma t}$ \\
\\
$\rho _{02,02}=-\frac{2}{3}e^{-3\Gamma t}+\frac{1}{3}e^{-4\Gamma t}+\frac{1}{%
3}e^{-2\Gamma t}$%
\end{tabular}%
\end{equation}%
are the density matrix elements $\rho _{ij,kl}.$ These eigenvalues~(\ref{eig}%
) are plotted in Fig.~\ref{Fig2}, where we observe that the times
where they vanish are in exact agreement with the times where
abrupt changes in the entanglement evolution appear. From
Eqs.~(\ref{eig}), these times can be analytically calculated in
terms of the parameter $\lambda $:
\begin{equation}
t_{1}=\ln(2/(2-\lambda)), t_{2}=\ln( 1/(1-\lambda)),
t_{3}=\ln(1/(1-\sqrt{\lambda })). \label{times}
\end{equation}%
Fig.~\ref{Fig3} shows the smooth behaviors of these times as a function of
the parameter $\lambda $ defining particular two qutrits mixed states. From
this picture we realize that the abrupt changes in the dynamics of
entanglement will appear for all values of $\lambda $ in the interval $[0,1]$%
. In particular for the maximally entangled state with $\lambda
=1$, there is sudden change for $t_{1}=\ln 2$, and the time of the
second and third sudden change, which is the ESD, goes to
infinite, showing that the entanglement decays asymptotically.
Note that this result differs substantially from its two-qubit
counterpart where the corresponding maximally entangled states
disentangle smoothly~\cite{Marcelo}. Also note that these abrupt
changes can be mathematically interpreted as discontinuities of
the derivative for the expression $\Lambda
(t)=\sum_{i=1}^{9}\sqrt{E_{i}\left( t\right) }$, and a sudden
change in the evolution of $\Lambda $ occurs whenever one of the
nine eigenvalues $E_{i}$ becomes zero, as observed in
Fig.~\ref{Fig2}

\begin{figure}[t]
\includegraphics[width=60mm]{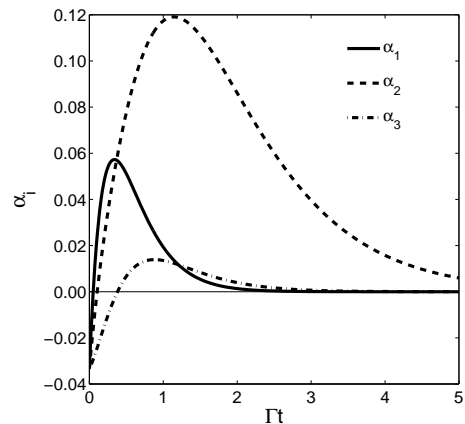}
\caption{{Relevant eigenvalues of partial transpose matrix $\protect\rho%
^{T_{A}}$.}}
\label{Fig4}
\end{figure}

The analysis to explain these particular sudden changes in the evolution of
entanglement has been done in terms of the eigenvalues of the matrix $M$.
However, we can also understand these abrupt changes in $\Lambda (t)$ by
observing the behavior of the eigenvalues of the partial transpose matrix $%
\rho ^{T_{A}}$. In our case only three eigenvalues give us
information about these sudden changes and are plotted in
Fig.~\ref{Fig4}. We notice that these eigenvalues change from
negative to positive values for specific times which are in
agreement with the sudden changes in the entanglement evolution.
In other words, the disentanglement rate changes whenever the rank
of the partially transposed matrix changes abruptly. We can also
associate to each eigenvalue of $\rho ^{T_{A}}$ a corresponding
entanglement witness operator such that $\alpha _{i}\left(
t\right) =Tr\left( W_{i}\rho \left( t\right) \right) $ with
$i=1,2,3$ and each $W_{i}$ is given by
\begin{equation}
\begin{tabular}{l}
$W_{1}=\frac{1}{2}\left[ |21\rangle \langle 21|-|11\rangle \langle
22|-|22\rangle \langle 11|+|12\rangle \langle 12|\right] ,$ \\
$W_{2}=\frac{1}{2}\left[ |10\rangle \langle 10|-|00\rangle \langle
11|-|11\rangle \langle 00|+|01\rangle \langle 01|\right] ,$ \\
$W_{3}=\frac{1}{2}\left[ |02\rangle \langle 02|-|00\rangle \langle
22|-|22\rangle \langle 00|+|20\rangle \langle 20|\right] .$%
\end{tabular}%
\end{equation}
At $t=0$, all three operators can be used to identify entanglement in $\rho$%
. As time goes by, they consecutively lose this capacity until there is no
entanglement left. This suggests a geometrical interpretation to the
phenomena here described which will be explored in further publications.

\begin{figure}[t]
\includegraphics[width=60mm]{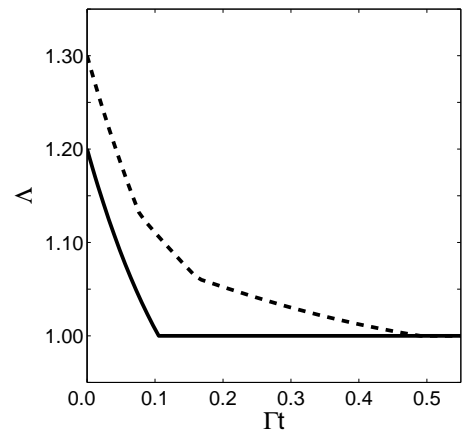}
\caption{{Entanglement evolution for the case shown in Fig
(\protect\ref {Fig1}) (dashed line) for $\protect\lambda =0.15$
and for a initial state in the subspace \{$\mid 11\rangle $, $\mid
22\rangle $\} (solid line) for $ \protect\chi =0.2$ }}
\label{Fig5}
\end{figure}

It is interesting to compare the case analyzed previously with
that of an initial state restricted to a two-dimensional subspace
$\rho=(1/2)(|11\rangle\langle 11|+|22\rangle\langle
22|+\chi|11\rangle \langle 22|+\chi|22 \rangle\langle 11|)$.
Fig.~\ref{Fig5} shows the evolution of the entanglement for
$\chi=0.2$ as compared with the state in Eq.~\ref{initial} for
$\lambda=0.15$. We observe that in the case of the initial
condition restricted to two dimensional subspace we have only one
abrupt change in the evolution corresponding to the ESD which
resembles the behavior observed for two qubits~\cite{Marcelo}. If
we look at the eigenvalues of $M$ we see that at the time when
$\Lambda $ goes to zero there is also an eigenvalue that goes to
zero, indicating that an abrupt change occurs. A similar
conclusion could actually be obtained when looking at the
negativity instead of Concurrence for the case of two qubits.

\begin{figure}[t]
\includegraphics[width=60mm]{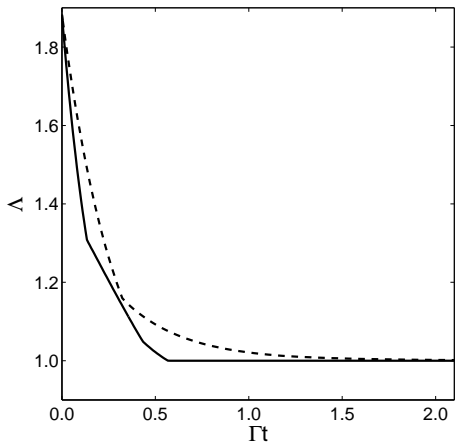}
\caption{{Entanglement evolution for the non maximal pure state.
Dashed line corresponds to $\protect\alpha =0.2386$, $\protect\beta =0.9545$, $%
\protect\gamma=0.1790$ and solid line corresponds to $\protect\alpha =0.1790$, $\protect\beta %
=0.2386$, $\protect\gamma=0.9545$. }}
\label{Fig6}
\end{figure}

In addition we could explore the entanglement evolution for
initial non maximally pure entangled states, for example, $\mid
\Phi \rangle =\alpha \mid 00\rangle +\beta \mid 11\rangle +\gamma
\mid 22\rangle $. For the sake of simplicity we consider $\alpha
$, $\beta $ and $\gamma $ real positive numbers. In this case, a
richer dynamics for the entanglement can be observed. Depending on
the choice of the amplitudes we can have asymptotic decay, sudden
death, sudden changes or a combination of them. Times
corresponding to the sudden changes and the ESD time are given by:
\begin{equation}
\begin{tabular}{l}
$t_{1}=-\frac{1}{\Gamma }\ln \left( 1-\frac{\beta }{2\gamma
}\right)$, $t_{3}=-\frac{1}{\Gamma }\ln \left( 1-\frac{\alpha
}{\gamma }\right) ,$ \\
$t_{2}=-\frac{1}{\Gamma }\ln \left[ 1-\frac{\beta }{3\gamma
}\left( 1+\frac{1}{\left( 2Z\right) ^{1/3}}-\left(
\frac{Z}{2}\right) ^{1/3}\right) \right] ,$
\end{tabular}
\end{equation}
where
\begin{equation}
Z=5-27\frac{\alpha \gamma }{\beta
^{2}}+3\sqrt{3}\left(1-10\frac{\alpha \gamma }{\beta
^{2}}+27\left( \frac{\alpha \gamma }{\beta ^{2}}\right)
^{2}\right)^{1/2}.
\end{equation}
From these expressions we realize that the entanglement dynamics
exhibit: $(a)$ asymptotic decay for ($\alpha \ge \beta > \gamma)$,
$(b)$ one sudden change and asymptotic decay for ($\beta \ge
\alpha \ge \gamma$, or $\alpha > \gamma>\beta )$, $(c)$ two sudden
changes and asymptotic decay  for ($\beta > \gamma > \alpha)$, and
$(d)$ two sudden changes and ESD for ($\gamma
> \beta > \alpha)$. Fig.~\ref{Fig6} shows two particular dynamics
evolution for the cases $(b)$ and $(d)$.

In summary we have studied the evolution of entanglement for high
dimensional dissipative quantum systems. By evaluating the
entanglement contained in the system using the Chen, Albeverio and
Fei measure we have observed outstanding new effects. Quantum
correlations undergo abrupt changes as precursors of ESD. These
can be characterized by observing the eigenvalues of the $M$
matrix which defines the amount of entanglement for the quantum
system. The dynamical changes are related to sudden changes in the
rank of the Matrix $M$. Similar behavior can be found for both
initially mixed or pure states and the ESD is recovered as a
particular case of these sudden dynamical changes.

FL and CEL acknowledge the financial support from MECESUP USA0108.
GR from CONICYT Ph. D. Programm Fellowships. MFS acknowledges
support from Mil\^ enio Infoquant/CNPq and thanks to Universidad de
Santaigo de Chile for the hospitality. JCR acknowledges support from
Fondecyt 1070157 and Milenio ICM P02-049.


\begin{thebibliography}{99}
\bibitem{nielsen} M.A. Nielsen and I.L. Chuang, \textit{Quantum Computation
and Quantum Information} (Cambridge Univ. Press., Cambridge, 2000).

\bibitem{Schroedinger} E. Schroedinger, Naturwissenschaften \textbf{23}, 807
(1935), Translation by John D. Trimmer. Published in Quantum Theory and
Measurement (J.A. Wheeler and W.H.Zureck, eds., Princeton University Press,
New Jersey 1983).

\bibitem{Diosi} L. Di\' osi, Lect. Notes Phys. 622, 157-163 (2003)

\bibitem{Dodd} P. J. Dodd and J. J. Halliwell, Phys. Rev. A, \textbf{69}, 052105 (2004)

\bibitem{eberly00} T. Yu and J. H. Eberly, Phys. Rev. Lett. \textbf{93},
140404 (2004); idem \textbf{97}, 140403 (2006).

\bibitem{eberly1} M. Y\"{o}na\c{c}, T. Yu and J. H. Eberly, J. Phys. B
\textbf{39}, S621 (2006).

\bibitem{Marcelo} M. Fran\c{c}a Santos, P. Milman, L. Davidovich, and N.
Zagury, Phys. Rev. A. \textbf{73}, 040305(R), 2006.

\bibitem{jamroz} A. Jamr\'{o}z, J. Phys. A \textbf{39}, 7727 (2006).

\bibitem{ficek} Z. Ficek and R. Tana\'{s}, Phys. Rev. A \textbf{74}, 024304
(2006).

\bibitem{Derkacz06} L. Derkacz and L. Jak\'{o}bczyk, Phys. Rev. A
\textbf{74}, 032313 (2006).

\bibitem{Albeverio} K. Chen, S. Albeverio, S. M. Fei, Phys. Rev. Lett. \textbf{
95}, 210501 (2005).

\bibitem{Peres} A. Peres, Phys. Rev. Lett. \textbf{77}, 1413 , 1996.

\bibitem{Horodecki} M. Horodecki, P. Horodecki, R. Horodecki, Phys. Lett. A
\textbf{223}, 1 (1996).

\bibitem{Rudolph} O. Rudolph, quant-ph/0202121.

\bibitem{Chen} K. Chen, L. A. Wu, Quantum Inf. Comput. 3, 193 (1999).
\end{thebibliography}
\end{document}